# Growth of Oriented Au Nanostructures: Role of Oxide at the Interface

A. Rath[1], J. K. Dash[1], R. R. Juluri[1], A. Rosenauer[2], Marcos Schoewalter[2], and P. V. Satyam[1,*]

[1] Institute of Physics, Sachivalaya Marg, Bhubaneswar - 751005, India

[2] Institute of Solid State Physics, University of Bremen, D-28359 Bremen, Germany

**Abstract**

We report on the formation of oriented gold nano structures on Si(100) substrate by annealing procedures in low vacuum ($\approx 10^{-2}$ mbar) and at high temperature ($\approx 975º$ C). Various thicknesses of gold films have been deposited with $SiO_x$ (using high vacuum thermal evaporation) and without $SiO_x$ (using molecular beam epitaxy) at the interface on Si(100). Electron microscopy measurements were performed to determine the morphology, orientation of the structures and the nature of oxide layer. Interfacial oxide layer, low vacuum and high temperature annealing conditions are found to be necessary to grow oriented gold structures. These gold structures can be transferred by simple scratching method.

*PACS:* 81.07.-b, 68.37.-d

*Keyword: Nanoscale Au structures, Surface and Interfaces, Electron Microscopy, Low Vacuum*

## 1. Introduction

Interactions taking place at the metal/semiconductor interface with or without an oxide layer at the interface is of great technological importance. It would be interesting to study these interfaces to understand their chemical and physical nature [1-3]. It is known that utility of such systems in microelectronics, gas sensors, etc., is greatly influenced by the interplay between strain energy and surface energy at surfaces and interfaces. The morphology (dimension, shape and number density) of 3D islands grown could be controlled by tuning the thickness and growth environment (for example temperature and vacuum)..

In recent years, Au nano/microstructures were studied actively due to their variety of applications in the emerging area of nanotechnology [4-6]. For example, the gold nanostructures have been used for tuning the surface plasmon resonance from the visible to the near IR range of the spectrum by changing the dimensions or shape of nanostructures[7]. In spite of generally being considered chemically inert in its bulk form, ultra-fine gold nanoparticles are known to be active catalysts (more so when supported by an oxide surface) [8]. As a result, various methods of preparation of gold nano/micro structures (spheres, rods, trapezoids, wires, triangles, squares etc) have been well reported [9-13]. Recently, Sanjeev *et al* have studied the structural evolution of Au films on Si (100) and Si (110) substrates in $N_2$ atmosphere and found that the resulting structures reorient themselves at higher temperatures [14]. Previous reports suggest the formation of faceted gold nanostructures, such as tetrahedrons, octahedrons, decahedrons and icosahedra structures. Surface energy anisotropy has been incorporated to understand the facet formation, crystal nucleation, growth and surface dynamics [15-20].

In this report, we discuss the influence of a native oxide and thermally grown oxide layer at the interface of Au/Si(100) systems, in the formation of highly oriented Au nanostructures at high temperatures. In this study, we have used a high vacuum (HV) and ultra high vacuum (UHV) – molecular beam epitaxy (MBE) facilities to grow various thicknesses of Au film on different surface conditions of the substrate such as with or without native/thermally grown oxide layer at the interface of gold and silicon substrate) .The resulting systems were characterized by various electron microscopy methods such as, field emission gun based scanning electron microscopy (FEGSEM) at low and high energies, (scanning) transmission electron microscopy ((S)TEM) in combination with electron energy loss spectroscopy (EELS).

## 2. Experimental

Gold thin films of varying thicknesses (5.0 nm, 11.7 nm and 50.0 nm) were deposited on Si(100) substrates while keeping the native oxide layer by thermal evaporation method under high vacuum conditions. To check the role f oxide layer thickness, an oxide layer of 50.0 nm thickness was thermally grown (using dry oxidation method) on Si(100) substrate and followed this by depositing a 50.0 nm Au thin film by HV thermal evaporation methods (hence forth these systems are denoted as Au/SiO$_x$/Si). To have a cleaner interface (i.e. without oxide at the interface), an ≈11.7 nm thick Au film was deposited on Si(100) under ultra high vacuum (UHV) conditions with a base pressure of $\approx 2 \times 10^{-10}$ mbar [11]. For this purpose, we have used molecular beam epitaxy (MBE) system where in the surface was cleaned by flashing the substrate at a temperature of about 1200ºC [11]. This procedure resulted in not having a native oxide present at the interface of Au and Si (hence forth these systems are denoted as Au/Si). The as-deposited Au/SiO$_x$/Si and Au/Si systems were annealed in the low vacuum furnace ($\approx 10^{-2}$ mbar) at $\approx 975^0$C for 30 minutes. Cross-sectional TEM (XTEM) specimens were prepared using mechanical thinning followed by low energy Ar$^+$ ion milling. For scanning transmission electron microscopy (STEM) based EELS measurements, specimens were prepared using focused ion beam. The TEM characterization of the above samples was done with 200 keV electrons (JEOL JEM 2010). SEM (scanning electron microscopy) measurements were carried out with 1.0 – 20 kV electrons (Neon40, Cross Beam, Carl Zeiss) and the scanning TEM (STEM) measurements were done using 300 keV electrons in the $C_s$-corrected FEI Titan 80/300 system.

**3. Results and discussions**

Fig 1(a) shows the bright field TEM image of as deposited 5.0 nm Au thin film on Si(100) substrate. The thin film seems to be continuous with almost 50% surface coverage. Fig. 1(b) depicts the SEM micrograph of the above sample which was thermally annealed at ≈ 975$^0$C under low vacuum condition. Formation of gold nanostructures could be seen with an average diameter of 17.3 ± 1.2 nm. Fig.1(c) shows the bright field XTEM image of the gold nano structures with an average height of 21.8 ± 1.1 nm. It is to be noted that the planar TEM images suggest the formation of spherical nano particles, where as the XTEM image reveals the beginning of the facet formation. To understand the role of thickness of the gold film, similar studies were carried out for 11.7 nm Au/SiO$_x$/Si(100) and 50nm Au/SiOx/Si(100) as well. As deposited 11.7 nm Au/SiO$_x$/Si(100) system shows almost a continuous layer with a surface coverage ~ 99% (Fig. 1 (d)). Annealing under low vacuum conditions for about 30 minutes leads to the formation of multi–faceted three-dimensional (3D) gold structures with an average size of 248 ± 13.1 nm (Fig 1(e)) and an average height of 210 ± 9.6 nm (Fig 1(f)). The as-deposited 50 nm/SiOx/Si(100) system shows almost a continuous layer (Fig.1(g)) and formation of well oriented faceted 3D gold structures with average size 530 ± 18.6 nm (fig 1(h)) and average height 291 ± 11.6 nm (fig 1 (i)) were observed upon annealing under the above mentioned condition. It is to be noted that, in both of the above systems height of gold particles is lower than their width. Fig. 1 (i) shows a cross-sectional TEM image of one of the faceted structures and the inset figure depicts selected area diffraction (SAD) taken on a single structure. The spots correspond to a d-spacing of 0.235 nm, 0.118 nm, 0.058 nm and 0.083 nm which matched with the (111), (222), (444) and (224) planes of Au [21]. It indicates that gold islands were crystalline in nature. The angle of contact between a typical faceted island and the substrate was found to be $\approx 126^0$. We have also calculated the angles between the faceted planes and almost all the faceted particles are bounded by {111}, {$\bar{1}$00} and {211} planes which match with the studied done by L.D. mark [17]. Due to high contact angle the faceted particles are easily scrubbed from the substrate. Our studies showed that top surfaces of all the particles exhibit well developed {111} facets parallel to the substrate surface. This can be explained by the fact that the surface energy of FCC metal has a minimum for {111} orientation [22].

SEM based elemental mapping (Fig. 2(a)-2(c)) shows the presence of Au in the faceted structure and Si from the substrate. Fig. 2(d) shows the high resolution SEM image (taken at 5keV electron energy), where facets can be clearly seen. To investigate the effect of thickness of the interfacial oxide, 50 nm SiOx layer was thermally grown on Si substrate and following this a 50 nm thick Au film (~50nm) was deposited under high vacuum conditions. This system, i.e., 50nm Au/50nm SiOx/Si(100) was subjected to low vacuum annealing for 30 minutes as explained earlier. Formation of 3D faceted islands has been observed, interestingly similar to earlier system (Fig 2(e)). This result showed that the thickness of the oxide layer does not seem to affect the facet. On the other hand, effect of the absence of any oxide layer at the Au/Si interface was demonstrated by carrying out similar studies on the third system, 11.7 nm Au/Si(100). Native oxide layer was removed at high temperature under UHV conditions (as explained in the experimental section) and a 11.7 nm Au thin film was deposited in the MBE system. This system was subjected to similar low vacuum annealing treatment and no formation of multifaceted 3D gold structures was observed (Fig. 2(f)). Instead, some fractal type of textures were formed along with irregular gold islands (Fig. 2(f)) indicating that the interfacial oxide plays an important role in the formation of faceted particles. The thickness of the oxide layer does not affect the facet formation but presence of oxide is necessary for the formation of facets. In MBE grown case, there is no oxide layer at the interface. Hence at higher temperature, both interfacial diffusion and the surface energy minimization processes occur. Due to the competition between the above two processes, equilibrium shape is not achieved by the system. Where as in oxide case, surface minimization process is dominant (no alloy formation has been seen) which leads to facet formation. The oxide at the interface inhibits the process of silicide at the Interface and helps in growing faceted Au gstructures.

In the following, we present results on the variation of the native oxide layer thickness at the Au and Si interface. Figure 3(a) depicts a STEM image taken from 50 nm Au/2 nm $SiO_x$/Si(100) annealed at ≈ 975 º C under low vacuum conditions. Fig. 3(b) shows one of the corners of the faceted structure at a higher magnification. EELS scans (integrated counts in EELS oxygen data) have been taken from region (i) and region (ii). The full-width at half maximum (FWHM) of the oxygen peak varies from 3.5 nm to 6.6 nm (Figs. 3 c (i) and d (ii)). This confirms the effective barrier role of an oxide layer inhibiting the Si out diffusion and Au inter-diffusion. The interface between (Fig. 3(a)) the Si substrate and the Au microstructures appears to be sharp without any Au diffusion into Si. But, for Au films grown under MBE conditions, gold inter-diffusion was found to be very prominent and this reduces the effect of later-diffusion of Au on the oxide surface and hence no oriented structures of Au are seen for the films that are grown under MBE conditions. The oriented structures (grown with oxide interface and low vacuum high temperature annealing procedures) were easily scratched with sharp blade and transferred to carbon coated grid to study further their structural and optical properties.

## 4. Conclusions

In summary, we reported a simple procedure to fabricate well oriented gold nanostructures by annealing Au/$SiO_x$/Si(100) systems under low vacuum and at high temperature. The shape of the oriented structures could be tuned by varying initial thickness of gold film: formation of faceted gold structures were observed in the case of higher gold thicknesses as opposed low thickness where formation of only spherical nanoparticles were seen. It was also shown that the oriented microstructures cannot be fabricated without oxide layer at the interface (using MBE method). Electron microscopy methods were used to determine the morphological and crystalline nature of these structures.

**Acknowledgements**

P V Satyam would like to thank the Department of Atomic Energy, Government of India, for 11th Plan Project No 11-R&D-IOP-5.09-0100 and the University of Bremen, Germany, for his sabbatical visiting program.

**Figure captions:**

**FIG. 1:** (a), (d) and (g) are SEM images of as-deposited samples of thickness 5.0 nm, 11.7 nm and 50.0 nm, respectively. (b), (e) and (h) are the SEM images taken at RT after heated at ≈ 975º C under low vacuum and corresponding XTEM images are shown in (c), (f) and (i). Inset of fig 1 (i) shows the SAD pattern taken on one of the faceted gold particle.

**FIG. 2:** (a) SEM image of 50nm – Au/SiO$_x$/Si(100) heated up to ≈ 975 º C and corresponding EDS (energy dispersive spectrometry) measurements taken from Si and Au secondary x-ray florescence in (b) for Si and (c) for Au. (d) SEM micrograph taken with 5.0 keV electrons (low-energy), (e) SEM image of 50 nm Au/thermally grown 50 nm SiOx/Si(100) substrate annealed at ≈ 975º C under low vacuum and (f) SEM of 11.7 nm Au/Si(100) (without native oxide) substrate annealed at ≈ 975º C under low vacuum conditions.

**FIG. 3:** (a) STEM image from 50 nm Au/SiOx/Si(100) annealed annealed at ≈ 975 º C. (b) is the region taken at higher magnification from a selected area in (a). Fig. c(i) is from the Oxygen EELS line scan (i) and d(ii) is from the region (ii) shown in (b). Width of the variation of oxide layer thickness can be seen in (c) and (d).

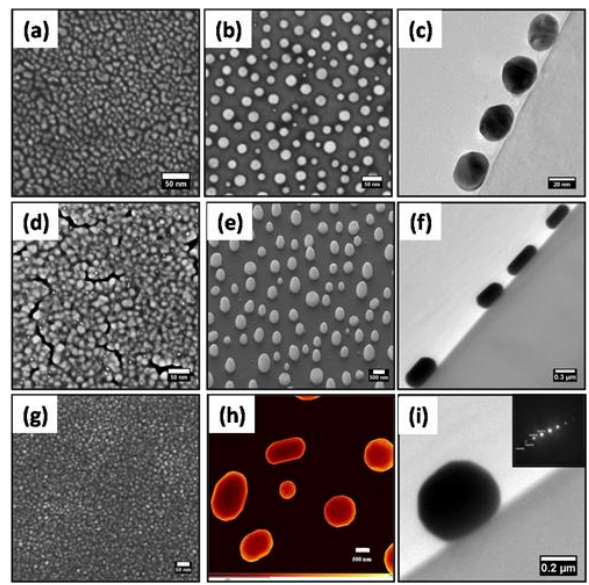

**FIG. 2: Rath et al.**

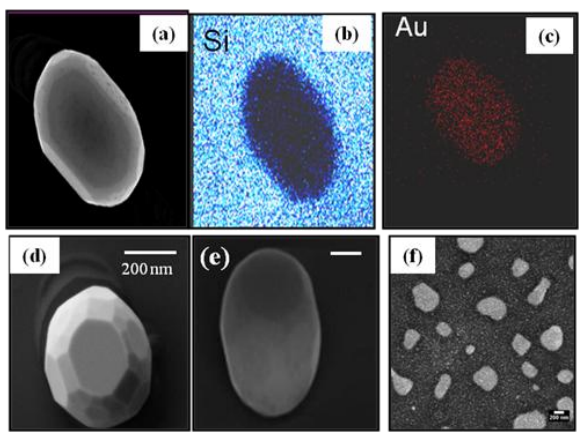

**FIG. 3: Rath et al.**

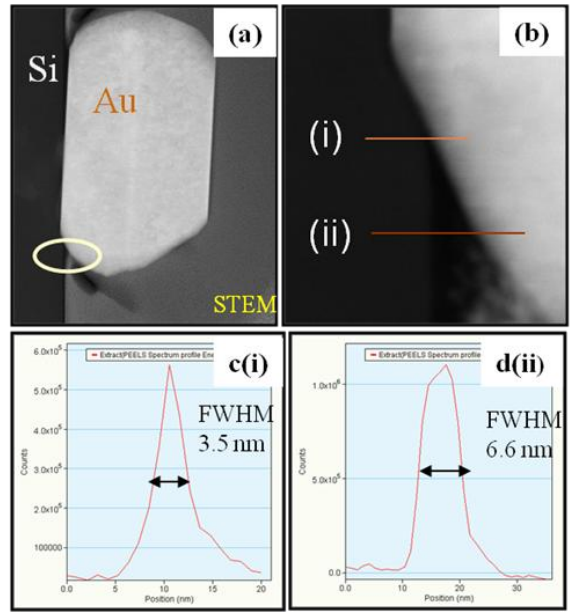

13